\documentstyle[12pt,psfig,epsf]{article}

\title{ TRANSVERSE MOMENTUM DEPENDENCE OF CUMULATIVE PIONS
\thanks{This work is supported by the Russian Foundation for Fundamental
Research, Grant No. 97-02-18123}
}
\author{
{\bf M.Braun, V.Vechernin}\\
{\it St. Petersburg State University, Russia}
}
\date{}

\def\bc{\begin{center}}
\def\ec{\end{center}}
\def\be{\begin{equation}}
\def\ee{\end{equation}}

\def\ub{\underbrace}

\def\del{\delta}
\def\lam{\lambda}
\def\Lam{\Lambda}

\def\olr{\ol{r}}

\def\ol#1{\overline{#1}}

\def\ran{\rangle}
\def\lan{\langle}

\def\lx1{x_1=\xi\rightarrow nx_0}

\def\hs#1{\hspace{#1cm}}

\def\v#1#2#3#4{
\put(#1,#2){\circle*{2.00}}
\put(#1,#3){\circle*{2.00}}
\put(#1,#2){\dashbox{1.5}(0,#4)[t]{}}
}

\begin{document}

\begin{titlepage}
\maketitle
\medskip

\begin{abstract}
In the framework of the recently proposed QCD based parton model for the
cumulative phenomena in the interactions with nuclei
the dependence of the cumulative pion  production rates on the
transverse momentum $K_{\perp}$ is studied.
The mean value of $K_{\perp}$ is found to grow with $x$ in the cumulative
region.
The obtained results are in agreement with the available experimental data.
\end{abstract}
\vspace*{1cm}
\end{titlepage}

\section{Introduction}

A few years ago we have proposed
a quark-parton model of cumulative phenomena
in the interactions with nuclei \cite{NPB94,YF97}
based on
perturbative QCD calculations of the corresponding quark
diagrams near the thresholds, at which other quarks
("donors") in the nuclear flucton transfer all their
longitudinal momenta to the distinguished active quark
and become soft.

Consider the scattering of
a hadronic projectile off a nucleus with the momentum $P$
in the c.m. system.
At high energies
the momentum $K$ of the produced pion belongs to
the cumulative region if $K_z>P_z/A$.
As a reasonable first approximation, we treat the nucleus as a
collection of $N=3A$ valence quarks, which, on the average, carry each
longitudinal momentum $x_{0}P_{z}/A$ with $x_{0}=1/3$.
In our approach the cumulative pion production proceeds in two steps.
First a valence quark with a scaling variable $x>1$ is created.
Afterwards
it decays into the observed hadron with its scaling variable $x$ close
to the initial cumulative quark's one. This second step is described by
the well-known quark fragmentation functions
\cite{CapellaT81} and will not be
discussed here.

The produced cumulative ("active") quark acquires
the momentum much greater than $x_{0}P_{z}/A$ only if this
quark has interacted by means of gluon exchanges
with other $p$ quarks of flucton ("donors")
and has taken some of their longitudinal momenta (see Fig.1).
If this active quark accumulates all
longitudinal momentum of these $p$ quarks then
$K_z=(p+1)x_0P_z/A$ and the donors  become soft.
It is well-known that interactions which make the longitudinal momentum
of one of the quark equal to zero may be treated by perturbation theory
\cite{Brodsky92}. This allows to calculate the part of Fig. 1
responsible for the creation of a cumulative quark explicitly.
This was done in \cite{NPB94,YF97}, to which papers we refer  the
reader for all the details. As a result we were able to explain the
exponential fall-off of the production rate in the cumulative region.

Since with  the rise of $x$ the active quark has to interact with
a greater number of donors, one expects that its average transverse momentum
also grows with $x$. Roughly one expects that $\langle K_{\perp}^2 \rangle$
is proportional to the number of interactions, that is, to $x$.
In \cite{NPB94,YF97}
this point was not studied:
we have limited ourselves with the inclusive cross-section
integrated over the transverse momenta,
which lead to some simplifications.
The aim of the present paper is to find
the pion production rate dependence on
the transverse momentum
and the mean value of the latter
as a function of $x$ in the cumulative region.
This dependence and also the magnitude of $\langle K_{\perp}^2 \rangle$
have been studied experimentally. The comparison of our predictions
with the data allows to obtain further support for our model and
fix one of the two its parameters (the infrared cutoff).

\section{The $K_{\perp}$ dependence}

Repeating the calculations of the diagram in Fig.1 described in
\cite{NPB94,YF97} but not
limiting  ourselves with the inclusive cross-section
integrated over the transverse momentum,
we readily find that all dependence upon
the transverse momentum $K_{\perp}$ of the produced particle
is concentrated in a factor:
\be
J (K_{\perp})=
\int
\rho_A(\ub{r,...,r}_{p+1}|\ub{\olr,...,\olr}_{p+1})
G(c_1,...,c_p)
\prod_{i=1}^{p}
\lam(c_i-r)\lam(c_i-\olr)d^2 c_i
e^{i(\olr-r)K_{\perp}}
d^2 r d^2 \olr
\ee
Here $\rho_A$ is the (translationally invariant)
quark density matrix of the nucleus:
\be
\rho_A(r_i|\olr_i)\equiv
\int
\psi_{\perp A}(r_i,r_m)
\psi^*_{\perp A}(\olr_i,r_m)
\prod_{m=p+2}^{N} d^2 r_m
\ee
where
$\psi_{\perp A}$ is the transverse part of
the nuclear quark wave function. The propgation of soft donor quarks
is decribed by
\be
\lam(c)=\frac{K_0(m|c|)}{2\pi}
\label{lam}
\ee
where $m$ is the constituent quark mass and  $K_0$
is the modified Bessel function (the Mac-Donald function).
The interaction with the projectile contributes a factor
\be
G(c_1,...,c_p)=
\int
\prod_{i=1}^{p}
\sigma_{qq}(c_i-b_i)
\eta_H(b_1,...,b_p) d^2 b_i
\ee
where $\sigma_{qq}(c)$ is the quark-quark cross-section
at a given value of impact parameter $c$ and
\be
\eta_H(b_1,...,b_p)=
\sum_{L\geq p}\frac{L!}{(L-p)!}
\int|\psi_{\perp H}(b_i)|^2
\del^{(2)}(\frac{1}{L}\sum_{i=1}^{L} b_i)
\ d^2 b_{p+1}...d^2 b_L
\ee
is a multiparton distribution in the projectile, expressed via
the transverse part of
its partonic wave function $\psi_{\perp H}$ .
If we integrate $J (K_{\perp})$ over $K_{\perp}$ we come back
to our old result (Eq. (33) in \cite{NPB94}):
$$
\int J (K_{\perp})
\frac{d^2 K_{\perp}}{(2\pi)^2}=
\rho_A(\ub{0,...,0}_{p+1}|\ub{0,...,0}_{p+1})
\int
G(c_1,...,c_p)
\prod_{i=1}^{p} \lam^2(c_i-r)
d^2 c_i d^2 r
$$

If one assumes factorization of the multiparton distribution
$\eta_H(b_1,...,b_p)$ then\\
$G(c_1,...,c_p)$ also factorizes:
\be
G(c_1,...,c_p)=
\prod_{i=1}^{p}
G_0(c_i)
\label{facG}
\ee
Following \cite{YF97} we use the quasi-eikonal approximation
for $\eta_H$:
$$
\eta_H(b_1,...,b_p)=
\xi^{(p-1)/2}\nu_H^{p}\prod_{i=1}^{p}\eta_H(b_i)
$$
where $\xi$ is the quasi-eikonal diffraction factor,
$\nu_H^{}$ is the mean number of partons in the projectile hadron
and the single parton distribution $\eta_H(b)$
is normalized to unity.
In a Gaussian approximation
for $\sigma(c)$ and $\eta_H(b)$ we find:
$$
G_0(c)=
\xi^{\frac{1}{2}-\frac{1}{2p}}
\frac{\nu_H\sigma_{qq}}{\pi r_{0H}^2}
e^{-\frac{c^2}{r_{0H}^2}}
$$
where $\sigma_{qq}$ is the total  quark-quark cross-section,
$r_{0H}^2=r_0^2+r_H^2$, $r_0$ and $r_H$ are the widths of
$\sigma(c)$ and $\eta_H(b)$ respectively.

With the factorised $G(c_1,...,c_p)$ (\ref{facG}) we have
$$
J (K_{\perp})=
\int
\rho_A(\ub{0,...,0}_{p+1}|\ub{\olr-r,...,\olr-r}_{p+1})
j^p(r,\olr)
e^{i(\olr-r)K_{\perp}}
d^2 r d^2 \olr
$$
where
$$
j(r,\olr)=
\int d^2 c
G_0(c)
\lam(c-r)
\lam(c-\olr)
$$
We also have used the translational invariance of the $\rho$-matrix.
Note that near the real threshold we have no spectators and
$$
\rho_A(\ub{0,...,0}_{p+1}|\ub{\olr-r,...,\olr-r}_{p+1})
=\rho_A(\ub{0,...,0}_{p+1}|\ub{0,...,0}_{p+1})
$$
In any case large $K_{\perp}$ corresponds to small $\olr-r$ so
we factor $\rho_A$ out of the integral in zero point.
In the rest integral we pass to the variables
$$
B=\frac{r+\olr}{2}, \hs 1 b=\olr-r
$$
and shift the integration variable $c$, then
\be
J (K_{\perp})=
\rho_A(\ub{0,...,0}_{p+1}|\ub{0,...,0}_{p+1})
\int
j^p(B,b)
e^{ibK_{\perp}}
d^2 b d^2 B
\ee
where
\be
j(B,b)=
\int
G_0(B+c)
\lam(\frac{b}{2}-c)
\lam(\frac{b}{2}+c)
d^2 c
\label{j}
\ee

\section{The calculation of $\lan |K_{\perp}|\ran $}

Now we would like to find the width of the distribution on $K_{\perp}$
as a function of $p$ or what is the same of the cumulative
number $x=(p+1)/3$.
From the mathematical point of view it is simpler to calculate
the mean squared width of the distribution $\lan K_{\perp}^2\ran $.
Unfortunately in our case this quantity is logarithmically
divergent at large $K_{\perp}$.
This divergency results from the behavior of $j(B,b)$
at small $b$. This behavior is determined by the behavior of the
$ \lam(b)=K_0(m|b|)/(2\pi) $ (\ref{lam}),
which has a logarithmical singularity
at $|b|=0$. Smooth $G_0(B+c)$ does not affect this behavior.

For this reason we shall rather calculate $\lan |K_{\perp}|\ran $:
\be
\lan |K_{\perp}|\ran
=\frac{1}{J_N}
\int
j^p(B,b)
|K_{\perp}|e^{ibK_{\perp}}
d^2 b d^2 B
\frac{d^2 K_{\perp}}{(2\pi)^2}
\ee
where $J_N$ is the same integral as in the numerator
but without $|K_{\perp}|$.
Presenting $|K_{\perp}|$ as $K_{\perp}^2/|K_{\perp}|$ and
$K_{\perp}^2$ as the Laplacian $\Delta_b$ applied to
the exponent we find
$$
\lan |K_{\perp}|\ran
=-\frac{1}{J_N}
\int
j^p(B,b)
\Delta_b e^{ibK_{\perp}}
d^2 b d^2 B
\frac{d^2 K_{\perp}}{|K_{\perp}|(2\pi)^2}
$$
Twice integrating by parts
and using the formula
$$
\int
\frac{d^2 K_{\perp}}{|K_{\perp}|}e^{ibK_{\perp}} =
\frac{2\pi}{|b|}
$$
we find
$$
\lan |K_{\perp}|\ran
=-\frac{1}{2\pi J_N}
\int
\frac{1}{|b|}
\Delta_b j^p(B,b)
d^2 b d^2 B
$$
Now we again integrate by parts once to find
$$
\lan |K_{\perp}|\ran
=-\frac{1}{2\pi J_N}
\int
d^2 B
\frac{d^2 b}{|b|^2}
(n_b \nabla_b) j^p(B,b)
$$
where  $n_b=b/|b|$. This leads to our final formula
\be
\lan |K_{\perp}|\ran
=-\frac{p}{2\pi J_N}
\int
d^2 B
\frac{d^2 b}{|b|^2}
j^{p-1}(B,b)
(n_b \nabla_b) j(B,b)
\label{Kperp}
\ee
where $j(B,b)$ is given by (\ref{j}),
$ \lam(b) $ is given by (\ref{lam}) and
$$
J_N=\int d^2 B j^{p}(B,b=0)
$$

\section{Approximations}
To simplify numerical calculations we make some additional
approximations, which are not very essential but are well supported
by the comparison with exact calulations at a few sample points.

As follows from the
the asymptotics of  $K_0(z)$ at large $z$
$$
K_0(z)\simeq\sqrt{\frac{\pi}{2z}}e^{-z}
$$
the width of $\lam(b)$ (\ref{lam}) is of the order $m^{-1}$.
The function $G_0$ is smooth in the vicinity of the origin
and its width $r_{0H}=\sqrt{r_0^2+r_H^2}$  is substancially
larger than the width of $\lam$.
For this reason we factor $G_0(B+c)$ out of the integral (\ref{j})
over $c$ at the point $B$:
\be
j(B,b)=
G_0(B)\Lam(b),
\hs 1
\Lam(b) \equiv
\int
\lam(c)
\lam(c-b)
d^2 c=
\frac{|b|}{4\pi m}K_1(m|b|)
\label{Lam}
\ee
Then we find that the integrals over $B$ and $b$ decouple
\be
J (K_{\perp})=
\rho_A(\ub{0,...,0}_{p+1}|\ub{0,...,0}_{p+1})
\int G^p_0(B) d^2 B
\int \Lam^p(b)
e^{ibK_{\perp}}
d^2 b
\ee

In this approximation we find that $\lan |K_{\perp}|\ran $
depends only
on one parameter - the constituent quark mass $m$,
which  in our approach plays the role of an infrared cutoff:
\be
\lan |K_{\perp}|\ran
=pm
\int_0^{\infty} dz K_0(z) (zK_1(z))^{p-1}
\label{apprKperp}
\ee
This allows to relate $m$ directly to the experimental data on
the transverse momentum dependence.

\section{Comparison with the data and discussion}

The integral in (\ref{apprKperp}) can be easy calculated
numerically. For values of $p=1,...,12$ it is very well
approximated by a power dependence (see Fig.2), so that we obtain
\be
\lan |K_{\perp}|\ran/m
=1.594\, p^{0.625}
\ee
As we observe, the rise of $\lan |K_{\perp}|\ran$ turns out to be
even faster than
expected on naive physical grounds mentioned in the Introduction
($\sim\sqrt{p}$).
The resulting plots for $\lan |K_{\perp}|\ran^2$
as a function of the cumulative number $x=(p+1)/3$
at different values of parameter $m$ are shown in Fig.3
together with avaiable experimental data from \cite{Boyarinov94}
on $\lan K_{\perp}^2\ran$ for pion production obtained in
experiments
\cite{Boyarinov94}-\cite{Boyarinov87}
with 10 $GeV$ protons
and \cite{Baldin82, Baldin83} with 8.94 $GeV$ protons.

Note that earlier publications of the first group
\cite{Boyarinov92,Boyarinov87}
reported a much stronger increase of $\lan K_{\perp}^2\ran$
with $x$, up to  value 2 $(GeV/c)^2$ at $x=3$ for pion production.
In our approach such an increase would require the quark mass to be as high
as $m \simeq 225 MeV$.
In a more recent publication \cite{Boyarinov94}
the rise of $\lan K_{\perp}^2\ran$ is substancially weaker
(it corresponds to $m \simeq 175 MeV$ in our approach).
The authors of \cite{Boyarinov94} explain this
by new experimental data obtained and by a cutoff $K_{\perp max}$
introduced in calculations of $\lan K_{\perp}^2\ran$ in \cite{Boyarinov94}.
The introduction of this cutoff considerably
(approximatly two times)
decreases the experimental value of $\lan K_{\perp}^2\ran$ at $x=3$.
In our opinion this is a confirmation that the
cumulative pion production rate only weakly decreases with $K_{\perp}$
in the cumulative region so that the
the integral over $K_{\perp}^2$ which enters the definition of
$\lan K_{\perp}^2\ran$ is weakly convergent or even divergent,
as in our approach.
Undoubtedly presentation of the experimental data in terms of
the mean value
$\lan |K_{\perp}| \ran^2$,
rather than $\lan K_{\perp}^2\ran$ should reduce
the dependence on the cutoff
$K_{\perp max}$ and make the results more informative.

One of the ideas behind the investigations of the cumulative
phenomena is that they may be a manifestation
of a cold quark-gluon plasma formed when several nucleons overlap in the
nuclear matter. In \cite{NPB94} we pointed out  that our model does not
correspond to this picture. It implies  coherent interactions of the
active quark with donors and, as a result, strong correlations between the
longitudinal and transverse motion. Predictions for the dependence of
$\lan |K_{\perp}| \ran$ on $x$ are also  different. From the cold
quark-gluon plasma model one expects $\lan |K_{\perp}| \ran$ to behave
as $x^{1/3}$, since the Fermi momentum of the quarks inside the overlap
volume is proportional to the cubic root of the quark density. Our model
predicts a much faster increase, with a power twice larger. The
experimental data seem to support our predictions.

\section{Acknowledgments}

The authors are greatly thankful to Prof. P.Hoyer who attracted
their attention to the problem.

This work is supported by the Russian Foundation for Fundamental
Research, Grant No. 97-02-18123.

\newpage

{\large {\bf Figure captions}}

\begin{description}

\item[Fig.~1]
The diagram for the production of a cumulative quark with the momentum $K$
in the scattering
of a projectile hadron with the momentum $H$
off a nucleus $A$ with the momentum $P$.
Dashed and chain lines show gluon and pomeron
exchanges, respectively.

\item[Fig.~2]
The $\lan |K_{\perp}|\ran/m$ as a function of $p$.
The points are the results of calculations on (\ref{apprKperp}).
The line is the best power fit.

\item[Fig.~3]
The $\lan |K_{\perp}|\ran^2$
as a function of the cumulative number $x=(p+1)/3$.
The lines are the results of calculations on (\ref{apprKperp})
at different values of parameter $m$.
The points ($\bullet$) are the experimental data
from \cite{Boyarinov94} on $\lan K_{\perp}^2\ran$ for pion production
with a cutoff (see the text)
obtained in experiments \cite{Boyarinov94}-\cite{Baldin83}
on a bombardment of nuclei by 10 $GeV$ and 9 $GeV$ protons.
The points ($\circ$) are the data from the earlier publications
of the group \cite{Boyarinov92,Boyarinov87} without a cutoff.

\end{description}

\vfill
\bc
M.Braun and V.Vechernin
\ec

\newpage
\thispagestyle{empty}
\vspace*{5cm}
\unitlength=0.5mm
\linethickness{0.4pt}
\noindent
\begin{picture}(70,140)(-80,-20)

\def\fs1{
\v{30}{40}{60}{20}
\v{40}{50}{60}{10}
\v{-30}{40}{60}{20}
\v{-40}{50}{60}{10}

\put(-50,10){\line(1,0){100}}
\put(-50,20){\line(1,0){100}}
\put(-50,30){\line(1,0){100}}
\put(-50,40){\line(1,0){100}}
\put(-50,50){\line(1,0){100}}

\put(55,35){\oval(10,60)[]}
\put(60,34){\line(1,0){10}}
\put(60,36){\line(1,0){10}}
\put(-55,35){\oval(10,60)[]}
\put(-70,34){\line(1,0){10}}
\put(-70,36){\line(1,0){10}}

\put(55,35){\makebox(0,0)[cc]{$A$}}
\put(-55,35){\makebox(0,0)[cc]{$A$}}
\put(75,35){\makebox(0,0)[lc]{$P$}}
\put(-75,35){\makebox(0,0)[rc]{$P$}}
}

\put(70,40){
\begin{picture}(70,150)

\fs1

\put(-50,60){\line(1,0){40}}
\put(10,60){\line(1,0){40}}

\multiput(-5,-10)(0,3){20}{\makebox(0,0)[cc]{$\circ$}}
\multiput(5,-20)(0,3){20}{\makebox(0,0)[cc]{$\circ$}}
\put(-5,-10){\circle*{4}}
\put(-5,50){\circle*{4}}
\put(5,-20){\circle*{4}}
\put(5,40){\circle*{4}}

\put(25,-20){\oval(10,30)[]}
\put(-25,-20){\oval(10,30)[]}
\linethickness{2.pt}
\put(30,-20){\line(1,0){20}}
\put(-30,-20){\line(-1,0){20}}
\linethickness{0.4pt}
\put(55,-20){\makebox(0,0)[lc]{$H$}}
\put(-55,-20){\makebox(0,0)[rc]{$H$}}

\put(-20,-10){\line(1,0){40}}
\put(-20,-20){\line(1,0){40}}
\put(-20,-30){\line(1,0){40}}

\put(15,62){\makebox(0,0)[cb]{$K$}}
\put(-15,62){\makebox(0,0)[cb]{$K$}}

\end{picture}
}
\end{picture}

\vfill
\bc
Fig.~1\\
M.Braun and V.Vechernin
\ec

\newpage
\thispagestyle{empty}
\ \
\vspace*{5cm}

\begin{figure}[h]
\centerline{\psfig{figure=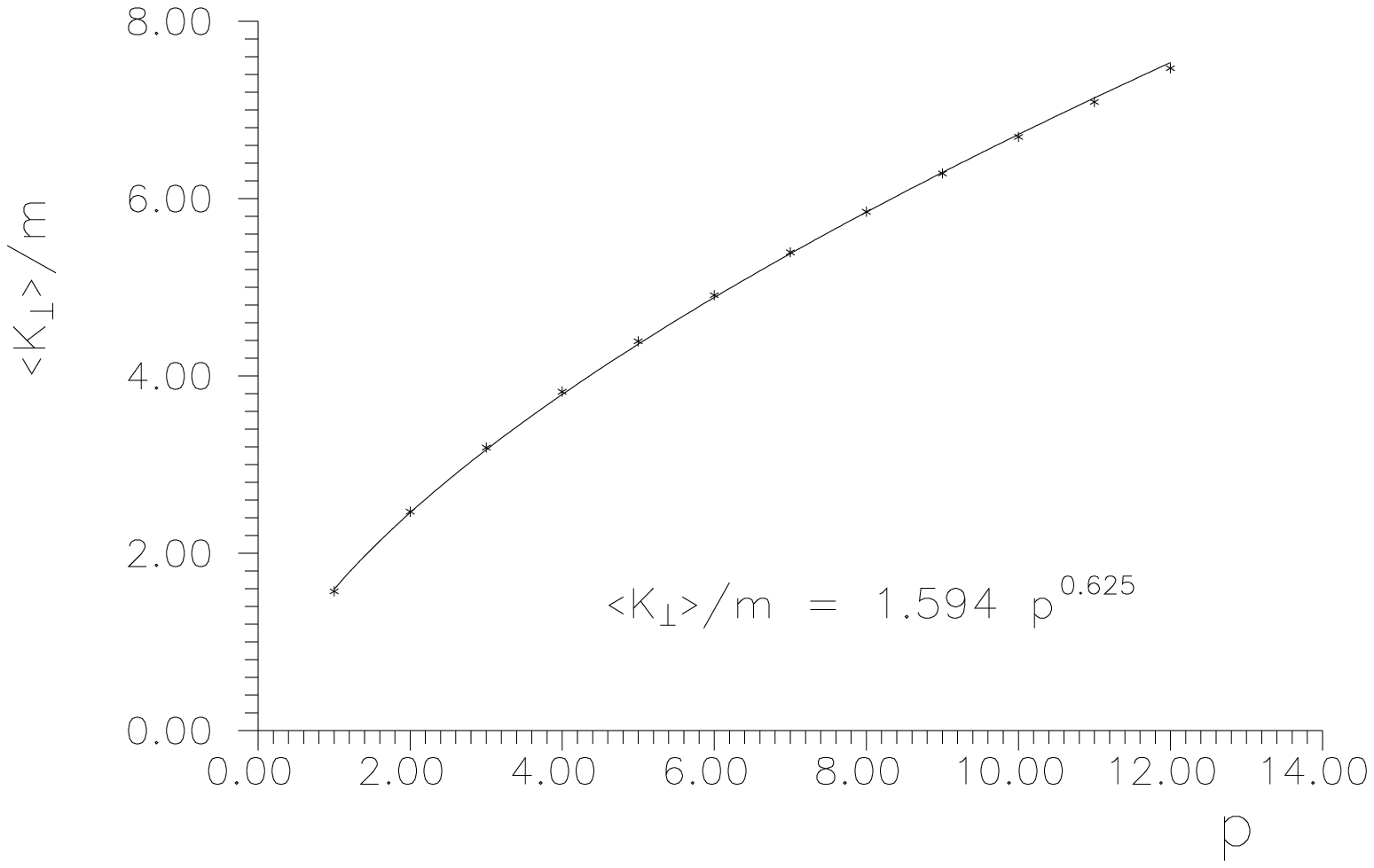}}
\end{figure}

\vfill
\bc
Fig.~2\\
M.Braun and V.Vechernin
\ec

\newpage
\thispagestyle{empty}
\ \
\vspace*{5cm}

\begin{figure}[h]
\centerline{\psfig{figure=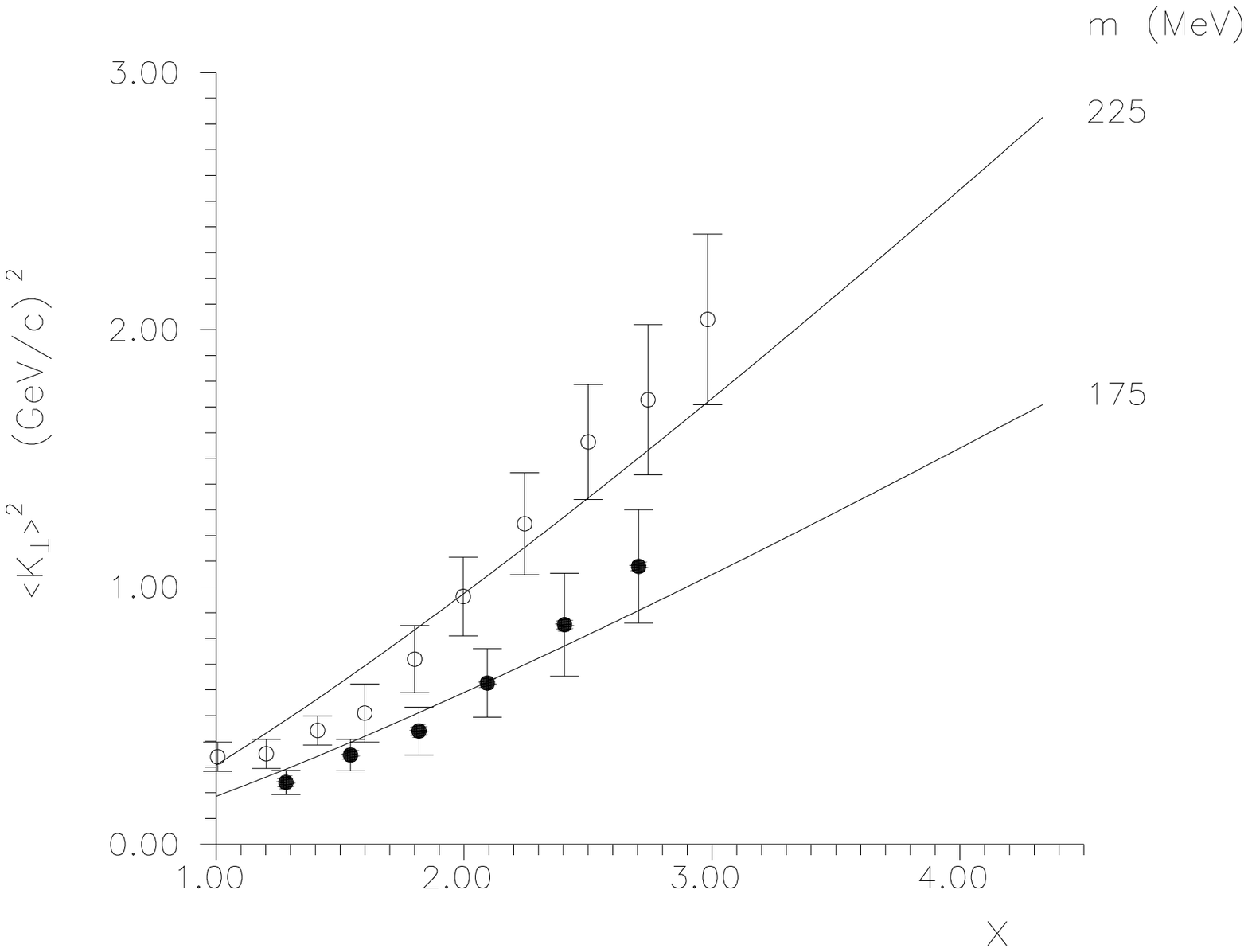}}
\end{figure}

\vfill
\bc
Fig.~3\\
M.Braun and V.Vechernin
\ec

\end{document}